\documentclass[%
 reprint,
superscriptaddress,
nofootinbib,
 amsmath,amssymb,
 aps,
prd,
]{revtex4-1}

\usepackage{graphicx}
\usepackage{dcolumn}
\usepackage{bm}
\usepackage{hyperref}
\usepackage{float}

\begin{document}

\preprint{APS/123-QED}

\title{Cosmological Parameter Biases from Doppler-Shifted Weak Lensing \\ in Stage IV Experiments}

\author{Anurag C. Deshpande}
\email{anurag.deshpande.18@ucl.ac.uk}
\affiliation{%
Mullard Space Science Laboratory, University College London, Holmbury St. Mary, Dorking, Surrey, RH5 6NT, UK}
\author{Thomas D. Kitching}%
\affiliation{%
Mullard Space Science Laboratory, University College London, Holmbury St. Mary, Dorking, Surrey, RH5 6NT, UK}%

\date{\today}

\begin{abstract}
The advent of Stage IV weak lensing surveys will open up a new era in precision cosmology. These experiments will offer more than an order-of-magnitude leap in precision over existing surveys, and we must ensure that the accuracy of our theory matches this. Accordingly, it is necessary to explicitly evaluate the impact of the theoretical assumptions made in current analyses on upcoming surveys. One effect typically neglected in present analyses is the Doppler-shift of the measured source comoving distances. Using Fisher matrices, we calculate the biases on the cosmological parameter values inferred from a \emph{Euclid}-like survey, if the correction for this Doppler-shift is omitted. We find that this Doppler-shift can be safely neglected for Stage IV surveys. The code used in this investigation is made publicly available\footnote{\url{https://github.com/desh1701/k-cut_reduced_shear}}.
\end{abstract}

\maketitle


\section{\label{sec:intro}Introduction}

The change in the observed shape of distant galaxies due to weak gravitational lensing by the large-scale-structure of the Universe (LSS), known as cosmic shear, is a powerful tool for performing precision cosmology. It is a particularly strong probe of dark energy \cite{DETFrep}. Existing cosmic shear surveys \cite{cfhtmain, DESpap, kids1000} are able to carry out cosmology competitive with modern Cosmic Microwave Background surveys \cite{Planck18}. The advent of Stage IV \cite{DETFrep} weak lensing surveys, like \emph{Euclid}\footnote{\url{https://www.euclid-ec.org/}} \cite{EuclidRB}, the \emph{Nancy Grace Roman Space Telescope}\footnote{\url{https://roman.gsfc.nasa.gov/}} \cite{WFIRSTpap}, and the Rubin Observatory\footnote{\url{https://www.lsst.org/}} \cite{LSSTpap}, will mean more than an order-of-magnitude increase in precision over the present generation of surveys.

In order to match this increased precision in the data, we must ensure that our theoretical analyses are sufficiently accurate. Accordingly, the impact of neglecting higher--order systematic effects on Stage IV experiments must be explicitly evaluated. In this work, we use the Fisher matrix formalism to predict the cosmological parameter biases from a \emph{Euclid}-like survey, when one such effect is neglected: the Doppler-shift of measured source redshifts due to their peculiar velocities and the inhomogeniety of the Universe. While the second--order correction for this effect is known \cite{DopplerBernardeau, CLSch18}, its impact at the angular power spectrum level, on intrinsic alignments (IA), and on cosmological parameter inference for the specifications of an \emph{Euclid}-like survey \cite{ISTFpap}, and under the Limber approximation, has not been explicitly evaluated.

This work is organised in the following manner: in Section \ref{sec:theory}, we detail our theoretical formalism. Here, the standard cosmic shear power spectrum calculation is described. We also review how contributions to the shear signal from non-cosmological IAs and shot noise are accounted for. Then, the procedure for correcting for the Doppler-shift of source redshifts when observing gravitational lensing is described. Our formalism for predicting cosmological parameter constrains and any biases in them in the presence of systematic effects, using Fisher matrices, is then explained. Following this, in Section \ref{sec:method}, we detail our method for modelling a Stage IV survey, and the fiducial cosmology chosen. Finally, in Section \ref{sec:results}, the results of our investigation are shown. We present the magnitude of the Doppler-shift correction to the angular power spectra relative to the angular power spectra themselves. Lastly, we state the cosmological parameter biases resulting from neglecting this correction, for a \emph{Euclid}-like survey.

\section{\label{sec:theory}Theory}

We begin this Section by describing the first--order cosmic shear power spectrum calculation. Next, we review the theoretical expressions for the contributions to the observed shear power spectra resulting from IAs and shot noise. Then, we outline the second--order correction to the cosmic shear angular power spectra that results from the Doppler-shift of source redshifts. Finally, we explain our use of Fisher matrices to predict cosmological parameter constraints and biases.

\subsection{\label{subsec:angpowfo}The First-order Cosmic Shear Power Spectrum}

The observed ellipticity of distant galaxies is distorted due to the weak gravitational lensing of their light by the LSS. This change in ellipticity is related to the reduced shear, $g$, which is given by:
\begin{equation}
    \label{eq:redshear}
    g^\alpha(\boldsymbol{\theta})= \frac{\gamma^\alpha(\boldsymbol{\theta})}{1-\kappa(\boldsymbol{\theta})},
\end{equation}
where $\boldsymbol{\theta}$ is the position of the source on the sky, $\gamma$ is the spin-2 shear with index $\alpha = 1, 2$, and $\kappa$ is the convergence. The shear encodes the part of weak lensing which results in an anisotropic stretching of the source image that would make a circular light distribution elliptical. Meanwhile, convergence is the component of weak lensing that causes an isotropic increase or decrease in the image's size. In the case of weak lensing, $|\kappa|\ll 1$, which allows us to make the reduced shear approximation:
\begin{equation}
    \label{eq:RSA}
    g^\alpha(\boldsymbol{\theta}) \approx \gamma^\alpha(\boldsymbol{\theta}).
\end{equation}
While this approximation results in significant cosmological parameter biases for Stage IV experiments \cite{Deshpap}, it has been shown that these can be sufficiently mitigated through the use of scale-cut techniques such as $k$-cut cosmic shear \cite{kcutrs}. Accordingly, we proceed under the reduced shear approximation for the remainder of this work.

For a tomographic redshift bin $i$, the convergence in its most general form, in spherical harmonic space, and on the celestial sphere, takes the form:
\begin{align}
    \label{eq:convspehe}
    \widetilde{\kappa}_{i; \ell m} &= 4\pi i^\ell \int_0^{\chi_{\rm lim}} {\rm d}\chi W_i(\chi) \int_0^\infty \frac{{\rm d}^3 k}{(2\pi)^3}j_\ell(k\chi)\nonumber\\
    &\times {}_2Y^*_{\ell m}(\boldsymbol{\hat{k}})\widetilde{\delta}(\boldsymbol{k}, \chi),
\end{align}
where $\ell = |\boldsymbol{\ell}|$ is the amplitude of the spherical harmonic conjugate of $\boldsymbol{\theta}$, $\chi$ is the comoving distance, $\chi_{\rm lim}$ is the limiting comoving distance of the survey, $j_\ell$ are spherical Bessel functions, ${}_2Y^*_{\ell m}$ are spin-weighted spherical harmonics with spin$=2$, $\widetilde{\delta}$ is the matter density contrast of the Universe, and $\boldsymbol{k}$ is a spatial momentum vector with magnitude $k=|\boldsymbol{k}|$. The convergence is a projection of the matter density contrast along the line-of-sight. Under the Limber approximation \cite{ExtendedLimber}, in which we consider only wave-modes in the plane of the sky to be contributing to the lensing signal, this simplifies to:
\begin{equation}
    \label{eq:convergence}
    \widetilde{\kappa}_{i}(\boldsymbol{\ell})=\int_{0}^{\chi_{\rm lim}} {\rm d}\chi\:\widetilde{\delta}(k, \chi)\:W_i(\chi),
\end{equation}
where now the vector $\boldsymbol{\ell}$ has angular component $\phi_\ell$, and magnitude $\ell$, and $k = (\ell + 1/2)/S_{\rm K}(\chi)$ \cite{ExtendedLimber}, with $S_{\rm K}(\chi)$ being a function that encodes the Universe's curvature, $K$. This is defined as:
\begin{equation}
    \label{eq:SK}
    S_{\rm K}(\chi) = \begin{cases}
    |K|^{-1/2}\sin(|K|^{-1/2}\chi) & \text{\small{$K>0$ (Closed)}}\\
    \chi & \text{\small{$K=0$ (Flat)}}\\
    |K|^{-1/2}\sinh(|K|^{-1/2}\chi) & \text{\small{$K<0$ (Open)}.}
  \end{cases}
\end{equation}
Additionally, $W_i$ is the lensing window function for tomographic bin $i$. This is given by the expression:
\begin{align}
    \label{eq:Wi}
    W_i(\chi) &= \frac{3}{2}\Omega_{\rm m}\frac{H_0^2}{c^2}\frac{S_{\rm K}(\chi)}{a(\chi)}\int_{\chi}^{\chi_{\rm lim}}{\rm d}\chi'\:n_i(\chi')\nonumber\\
    &\times\frac{S_{\rm K}(\chi'-\chi)}{S_{\rm K}(\chi')},
\end{align}
where $\Omega_{\rm m}$ is the dimensionless matter density of the Universe at present-day, $H_0$ is the Hubble constant, $c$ is the speed of light in a vacuum, $a(\chi)$ is the scale factor, and $n_i(\chi)$ is the galaxy probability distribution for bin $i$.

Making the flat-sky and `prefactor-unity' approximations \cite{limitsofshear17}, the relationship between shear and convergence is given by:
\begin{equation}
    \label{eq:fourier}
    \widetilde{\gamma}_i^\alpha(\boldsymbol{\ell})= T^\alpha(\boldsymbol{\ell})\,\widetilde{\kappa}_i(\boldsymbol{\ell}).
\end{equation}
Here, $T^\alpha(\boldsymbol{\ell})$ are trigonometric weighting functions:
\begin{align}
    \label{eq:Trigfunc1}
    T^1(\boldsymbol{\ell}) &= \cos(2\phi_\ell),\\
    \label{eq:Trigfunc2}
    T^2(\boldsymbol{\ell}) &= \sin(2\phi_\ell).
\end{align}

Considering an arbitrary shear field, we note two linear combinations of the individual shear components -- a curl-free $E$-mode, and a divergence-free $B$-mode:
\begin{align}
    \label{eq:Emode}
    \widetilde{E}_i(\boldsymbol{\ell})&=\sum_\alpha T^\alpha\:\widetilde{\gamma}_i^\alpha(\boldsymbol{\ell}),\\
    \label{eq:Bmode}
    \widetilde{B}_i(\boldsymbol{\ell})&=\sum_\alpha \sum_\beta \varepsilon^{\alpha\beta}\,T^\alpha(\boldsymbol{\ell})\:\widetilde{\gamma}_i^\beta(\boldsymbol{\ell}),
\end{align}
where $\varepsilon^{\alpha\beta}$ is the Levi-Civita symbol in two dimensions. When there are no systematic effects, the $B$-mode is zero, leaving the only the $E$-mode. From this, we define auto and cross-correlation angular power spectra, $C_{\ell;ij}^{\gamma\gamma}$:
\begin{equation}
    \label{eq:powerspecdef}
    \left<\widetilde{E}_i(\boldsymbol{\ell})\widetilde{E}_j(\boldsymbol{\ell'})\right> = (2\pi)^2\,\delta_{\rm D}^2(\boldsymbol{\ell}+\boldsymbol{\ell'})\,C_{\ell;ij}^{\gamma\gamma},
\end{equation}
where $\delta_{\rm D}^2$ is the two-dimensional Dirac delta function. These angular power spectra are defined as:
\begin{equation}
    \label{eq:Cl}
    C_{\ell;ij}^{\gamma\gamma} = \int_0^{\chi_{\rm lim}}{\rm d}\chi\frac{W_i(\chi)W_j(\chi)}{S^{\,2}_{\rm K}(\chi)}P_{\delta\delta}(k, \chi),
\end{equation}
where $P_{\delta\delta}(k, \chi)$ is the power spectrum of the matter density contrast. For a detailed overview of this calculation, see \cite{Kilbinger15}.

\subsection{\label{subsec:IAs}Intrinsic Alignments and Shot Noise}

In practice, the shear signal measured from surveys of galaxies contains not only the desired signal from cosmic shear, but other contributions as well. One of these non-cosmological contributions comes from the intrinsic alignment (IA) of galaxies \cite{JoachimiIAs}, as galaxies can have preferred, intrinsically correlated, alignments due to having formed in the same tidal environments. Taking into account this IA, to the first-order, the observed ellipticity of a galaxy, $\epsilon$, is expressed as:
\begin{equation}
    \label{eq:galelip}
    \epsilon = \gamma + \gamma^{\rm I} + \epsilon^s,
\end{equation}
where $\gamma$ is the cosmic shear due to the LSS, $\gamma^{\rm I}$ is the distortion from IAs, and $\epsilon^s$ is the ellipticity that the galaxy would have if no IA or cosmic shear was present. When we then construct a two-point statistic (e.g. the angular power spectrum) from this ellipticity, we find it has contributions from four types of terms: $\langle\gamma\gamma\rangle,\langle\gamma^{\rm I}\gamma\rangle$, $\langle\gamma^{\rm I}\gamma^{\rm I}\rangle$, and a shape (shot) noise component resulting from $\epsilon^s$.

Therefore, the observed angular power spectra, $C_{\ell;ij}^{\epsilon\epsilon}$, is the sum of each of these:
\begin{equation}
    \label{eq:ObsCl}
    C_{\ell;ij}^{\epsilon\epsilon} = C_{\ell;ij}^{\gamma\gamma} + C_{\ell;ij}^{{\rm I}\gamma} + C_{\ell;ij}^{\gamma{\rm I}} + C_{\ell;ij}^{\rm II} + N_{\ell;ij}^\epsilon,
\end{equation}
where $C_{\ell;ij}^{{\rm I}\gamma}$ are the correlation spectra between the background shear and the foreground IA, $C_{\ell;ij}^{\gamma{\rm I}}$ describe the correlation of the foreground shear with background IA which are zero except when photometric redshift estimates result in the observed redshifts being scattered between bins, $C_{\ell;ij}^{\rm II}$ are the IA auto-correlation spectra, and $N_{\ell;ij}^\epsilon$ encodes the shot noise.

We use the non-linear alignment (NLA) model \citep{NLAmodel} in order to describe the non-zero IA spectra:
\begin{align}
    \label{eq:cllig}
    C_{\ell;ij}^{{\rm I}\gamma} &= \int_0^{\chi_{\rm lim}}\frac{{\rm d}\chi}{S^{\,2}_{\rm K}(\chi)}[W_i(\chi)n_j(\chi)+n_i(\chi)W_j(\chi)]\nonumber\\
    &\times P_{\delta {\rm I}}(k, \chi),\\
    \label{eq:clli}
    C_{\ell;ij}^{\rm II} &= \int_0^{\chi_{\rm lim}}\frac{{\rm d}\chi}{S^{\,2}_{\rm K}(\chi)}n_i(\chi)n_j(\chi)\,P_{\rm II}(k, \chi),
\end{align}
where analogously to the shear angular power spectra, the IA angular power spectra are projections of the IA power spectra $P_{\delta {\rm I}}(k, \chi)$ and $P_{\rm II}(k, \chi)$. In the NLA model, these are proportional to the matter power spectrum:
\begin{align}
    \label{eq:pdi}
    P_{\delta {\rm I}}(k, \chi) &= \bigg[-\frac{\mathcal{A}_{\rm IA}\mathcal{C}_{\rm IA}\Omega_{\rm m}}{D(\chi)}\bigg]\:\:P_{\delta\delta}(k,\chi),\\
    \label{eq:pii}
    P_{\rm II}(k, \chi) &= \bigg[-\frac{\mathcal{A}_{\rm IA}\mathcal{C}_{\rm IA}\Omega_{\rm m}}{D(\chi)}\bigg]^2P_{\delta\delta}(k,\chi),
\end{align}
with $\mathcal{A}_{\rm IA}$ and $\mathcal{C}_{\rm IA}$ being free model parameters which are obtained through fitting to simulations or data, and $D(\chi)$ describing the evolution of the growth factor of density perturbations with comoving distance.

The final term in equation (\ref{eq:ObsCl}), which represents the shape (shot) noise, takes the form:
\begin{align}
    \label{eq:shotnoise}
    N_{\ell;ij}^\epsilon = \frac{\sigma_\epsilon^2}{\bar{n}_{\rm g}/N_{\rm bin}}\delta_{ij}^{\rm K},
\end{align}
under the assumption that the tomographic bins in the survey being studied are equi-populated. Here, $\sigma_\epsilon^2$ denotes the variance of the observed ellipticities in the sample of galaxies, $\bar{n}_{\rm g}$ is the surface density of galaxies, and $N_{\rm bin}$ is the survey's number of tomographic bins. The Kronecker delta, $\delta_{ij}^{\rm K}$, encodes the fact that galaxies' ellipticities at different redshifts should not be correlated; meaning that, for cross-correlation spectra, the shot noise will vanish.

\subsection{\label{subsec:Doppcorr}Doppler-shifted Cosmic Shear}

When measuring the effect of weak lensing on a given source, we observe its redshift. However, the inhomogeneity of the Universe, and the presence of the LSS means that the source will have a peculiar velocity towards its local overdensity. Consequently, the measured redshift will be perturbed by Doppler-shift. At the second-order, this will result in a correction to the observed reduced shear due to the coupling between this redshift perturbation and the lenses. Under the reduced shear approximation, this is given by \cite{DopplerBernardeau}:
\begin{equation}
    \label{eq:doppg}
    g^\alpha(\boldsymbol{\theta}, \chi) = \gamma^\alpha(\boldsymbol{\theta}, \chi) + \delta g_z(\boldsymbol{\theta}, \chi),
\end{equation}
where $\delta g_z$ accounts for the perturbation of the observed redshift according to:
\begin{equation}
    \label{eq:dgz}
    \delta g_z(\boldsymbol{\theta}, \chi) = -\frac{\mathrm{d} \gamma^\alpha}{\mathrm{~d} \chi} \frac{\mathrm{d} \chi}{\mathrm{~d} z} \delta z.
\end{equation}
Now, $\delta z$ is the perturbation of the source redshift due to Doppler-shift. Expanding this expression explicitly, and neglecting the sub-dominant Sachs-Wolfe and integrated Sachs-Wolfe effects results in:
\begin{equation}
    \label{eq:full_dg}
    \delta g_z(\boldsymbol{\theta}, \chi) = \frac{c}{\chi^2 H(\chi) a(\chi)} \, \boldsymbol{n} \cdot \boldsymbol{v} \int_0^\chi {\rm d}\chi \, \partial^2 \Phi (\boldsymbol{\theta}, \chi),
\end{equation}
where $H(\chi)$ is the value of the Hubble function at source comoving distance $\chi$, $\boldsymbol{n}$ is the unit direction vector pointing from the source to the observer, $\boldsymbol{v}$ is the peculiar velocity of the source, and $\Phi$ is the gravitational potential. In fact, $\delta g_z$ is a two-point term, as $\boldsymbol{n}\cdot\boldsymbol{v}$ also depends on the matter density contrast (see e.g. Appendix B of \cite{BaconDopp}). Accordingly, we write equation (\ref{eq:full_dg}) as a combination of $\kappa^{\rm like}$ and $\gamma^{\rm like}$ terms:
\begin{equation}
    \label{eq:dg_twopoint}
     \delta g_z(\boldsymbol{\theta}, \chi) = \kappa^{\rm like}(\boldsymbol{\theta}, \chi)\, \gamma^{\rm like}(\boldsymbol{\theta}, \chi),
\end{equation}
with:
\begin{align}
    \label{eq:klikereal}
    \kappa^{\rm like}(\boldsymbol{\theta}, \chi) = \frac{c}{\chi^2 H(\chi) a(\chi)}\, \boldsymbol{n}\cdot\boldsymbol{v},\\
    \label{eq:glikereal}
    \gamma^{\rm like}(\boldsymbol{\theta}, \chi) = \int_0^\chi {\rm d}\chi \, \partial^2 \Phi (\boldsymbol{\theta}, \chi).
\end{align}
The Doppler correction is now expressed as a product between a shear-like term, $\gamma^{\rm like}$, and a convergence-like term, $\kappa^{\rm like}$, analogously to the way in which other two-point correction terms (e.g. the reduced shear and magnification bias corrections \cite{Deshpap}) are typically formulated.

When expanded fully, in spherical harmonic space, and for a given tomographic redshift bin $i$, these terms take the form:
\begin{align}
    \label{eq:kdoppsph}
    \widetilde{\kappa}^{\rm like}_{i; \ell m} &= 4\pi i^{\ell}c \int_0^{\chi_{\rm lim}}\frac{\rm d\chi}{\chi^2 H(\chi) a(\chi)} n_i(\chi) \nonumber\\
    &\int_0^{\infty} \frac{{\rm} d^3 k}{(2\pi)^3}\,  \frac{j^\prime_\ell(k\chi)}{k}\, { }_{2} Y_{\ell m}^{*}(\hat{\boldsymbol{k}}) \widetilde{\delta}(\boldsymbol{k}, \chi),\\
    \label{eq:gdoppsph}
    \widetilde{\gamma}^{\rm like}_{i; \ell m} &= 4\pi i^{\ell} \frac{3\Omega_{\rm m} H_0^2}{2c^2} \int_0^{\chi_{\rm lim}} {\rm d}\chi n_i(\chi) \nonumber\\
    &\int_0^\infty \frac{{\rm} d^3 k}{(2\pi)^3} \, j_\ell(k\chi) \, { }_{2} Y_{\ell m}^{*}(\hat{\boldsymbol{k}}) \widetilde{\delta}(\boldsymbol{k}, \chi).
\end{align}
In equation (\ref{eq:kdoppsph}), $j^\prime_\ell$ is the derivative of the spherical-Bessel function $j_\ell$ with respect to $k\chi$.

Constructing an expression for the angular power spectrum which takes into account the additional Doppler correction term, under the flat-sky, flat-Universe, and Limber approximations (see Appendix for a detailed derivation) recovers equation (\ref{eq:Cl}), plus an additional term:
\begin{align}
    \label{eq:dcl}
    \delta C^{\rm Doppler}_{\ell;ij} &= \int_0^\infty \frac{{\rm d}^2 \boldsymbol{\ell^\prime}}{(2\pi)^2} \cos (2\phi_{\ell^\prime} - 2\phi_{\ell}) \nonumber\\
    &\times B_{ij}^{\rm Doppler}(\boldsymbol{\ell}, \boldsymbol{\ell'}, \boldsymbol{-\ell-\ell'}),
\end{align}
where:
\begin{align}
    \label{eq:Bdopp}
    B_{ij}^{\rm Doppler}(\boldsymbol{\ell}, \boldsymbol{\ell'}, \boldsymbol{-\ell-\ell'}) &= \int_0^{\chi_{\rm lim}} \frac{{\rm d}\chi}{\chi^4} [W^{\kappa\nu}_i(\chi, \ell') \nonumber \\
    &\times W^{\gamma\nu}_i(\chi)W_j(\chi) + W^{\kappa\nu}_j(\chi, \ell') \nonumber \\
    &\times W^{\gamma\nu}_j(\chi)W_i(\chi)] \nonumber\\
    &\times B^{\delta\delta\delta}(\boldsymbol{k}, \boldsymbol{k'}, \boldsymbol{-k-k'}, \chi)
\end{align}
Here, $B^{\delta\delta\delta}$ is the bispectrum of the matter density contrast, and $W^{\kappa\nu}_i$ and $W^{\gamma\nu}_i$ are weight functions, analogous to the lensing kernel of equation (\ref{eq:Wi}), corresponding to $\widetilde{\kappa}^{\rm like}_{i; \ell m}$ and $\widetilde{\gamma}^{\rm like}_{i; \ell m}$, respectively. We define these weight functions as:
\begin{align}
    \label{eq:Wkdopp}
    W^{\kappa\nu}_i(\chi, \ell) &= \left[\frac{\ell}{(\ell + 1/2)^2} - \frac{1}{(\ell + 3/2)}\right] \nonumber \\
    &\times \frac{c}{\chi \, H(\chi) \, a(\chi)}\, n_i(\chi), \\
    \label{eq:Wgdopp}
    W^{\gamma\nu}_i(\chi) &= \frac{3\Omega_{\rm m} H_0^2}{2c^2}\, n_i(\chi).
\end{align}
If we now also consider contributions from IAs, there will be another correction term to the angular power spectrum resulting from the correlation between the Doppler-shift and IA terms. This new term takes the form:
\begin{align}
    \label{eq:dcl_dopIA}
    \delta C^{\rm Doppler-IA}_{\ell;ij} &= \int_0^\infty \frac{{\rm d}^2 \boldsymbol{\ell^\prime}}{(2\pi)^2} \cos (2\phi_{\ell^\prime} - 2\phi_{\ell}) \nonumber\\
    &\times B_{ij}^{\rm \nu I}(\boldsymbol{\ell}, \boldsymbol{\ell'}, \boldsymbol{-\ell-\ell'}),
\end{align}
where now we define:
\begin{align}
    \label{eq:B-D-IA}
    B_{ij}^{\rm \nu I}(\boldsymbol{\ell}, \boldsymbol{\ell'}, \boldsymbol{-\ell-\ell'}) &= \int_0^{\chi_{\rm lim}} \frac{{\rm d}\chi}{\chi^4} [W^{\kappa\nu}_i(\chi, \ell') \nonumber \\
    &\times W^{\gamma\nu}_i(\chi)\,n_j(\chi) + W^{\kappa\nu}_j(\chi, \ell') \nonumber \\
    &\times W^{\gamma\nu}_j(\chi)\,n_i(\chi)] \nonumber\\
    &\times B^{\delta\delta{\rm I}}(\boldsymbol{k}, \boldsymbol{k'}, \boldsymbol{-k-k'}, \chi).
\end{align}
In equation (\ref{eq:B-D-IA}), $B^{\delta\delta{\rm I}}$ is the matter-IA bispectrum. In order to calculate this, we apply the ansatz which extends the NLA model to the bispectrum case \cite{Deshpap}; giving the expression:
\begin{align}
    \label{eq:BdI}
     B_{\delta\delta {\rm I}}(\boldsymbol{k_1},\boldsymbol{k_2},\boldsymbol{k_3},\chi) &= 2F_2^{\rm eff}(\boldsymbol{k_1},\boldsymbol{k_2}) P_{{\rm I}\delta}(k_1, \chi)P_{\delta\delta}(k_2, \chi) \nonumber\\
    &+ 2F_2^{\rm eff}(\boldsymbol{k_2},\boldsymbol{k_3})P_{\delta\delta}(k_2, \chi)P_{\delta {\rm I}}(k_3, \chi)  \nonumber\\
    &+ 2F_2^{\rm eff}(\boldsymbol{k_1},\boldsymbol{k_3})P_{\delta {\rm I}}(k_1, \chi) \nonumber \\
    &\times P_{\delta\delta}(k_3, \chi),
\end{align}
where $F_2^{\rm eff}$ is a fitting function obtained from N-body simulations, given in \cite{ScocCouch}.
\subsection{\label{subsec:fisher} Fisher Matrix Formalism}

To predict the cosmological parameter constraints for a \emph{Euclid}-like survey, we make use of Fisher matrices \citep{Tegmark97}; which are the expectation of the Hessian of the likelihood. The Fisher matrix depends exclusively on the mean of the data vector and on the covariance of the data when the assumption of a Gaussian likelihood is made. For weak lensing, it has been shown that this assumption is safe \cite{LSSTnongauss, gausspeter}. Additionally, the shear field has a mean value of zero. Accordingly, and making the additional assumption of a Gaussian covariance, the particular Fisher matrix we use is given by:
\begin{align}
    \label{eq:fishshear}
    F_{\tau \zeta}&=f_{\mathrm{sky}} \sum_{\ell=\ell_{\min }}^{\ell_{\max }} \Delta \ell\left(\ell+\frac{1}{2}\right) \nonumber\\
    &\times\operatorname{tr}\left[\frac{\partial \boldsymbol{C}_{\ell}}{\partial \theta_{\tau}} {\boldsymbol{C}_{\ell}}^{-1} \frac{\partial \boldsymbol{C}_{\ell}}{\partial \theta_{\zeta}} {\boldsymbol{C}_{\ell}}^{-1}\right],
\end{align}
where $f_{\rm sky}$ is the fraction of sky observed, $\Delta \ell$ is the bandwidth of $\ell$-modes sampled, these blocks in $\ell$ are summed over, and $\tau$ and $\zeta$ denote the current parameters of interest, $\theta_\tau$ and $\theta_\zeta$. A more detailed calculation of this expression can be found in \cite{ISTFpap}. For a given parameter, we are then able to predict the uncertainty using the expression:
\begin{align}
    \label{eq:sigfish}
    \sigma_\tau = \sqrt{{F_{\tau\tau}}^{-1}}.
\end{align}

If we want to predict how biased cosmological parameter values will be when neglecting a systematic effect within the data, this can be achieved by extending the Fisher matrix calculation \cite{Fishbias}, such that:
\begin{align}
    \label{eq:bias}
    b\left(\theta_{\tau}\right) &=\sum_{\zeta}F^{-1}_{\tau\zeta} f_{\mathrm{sky}} \sum_{\ell} \Delta \ell\left(\ell+\frac{1}{2}\right) \nonumber\\
    &\times  \operatorname{tr}\left[\delta \boldsymbol{C}_{\ell} \, {\boldsymbol{C}_{\ell}}^{-1} \frac{\partial \boldsymbol{C}_{\ell}}{\partial \theta_{\zeta}} {\boldsymbol{C}_{\ell}}^{-1}\right],
\end{align}
where the matrix $\delta \boldsymbol{C}_\ell$ contains the value of the systematic effect correction for the spectra of each tomographic bin auto and cross-correlation at a given $\ell$. Here, the Doppler-shift and Doppler-IA corrections of equation (\ref{eq:dcl}) and equation (\ref{eq:dcl_dopIA}), form this matrix.

\section{\label{sec:method}Methodology}

We study the effect of neglecting Doppler-shift on upcoming weak lensing surveys by utilising the forecasting specifications of a \emph{Euclid}-like survey \cite{ISTFpap} to represent Stage IV cosmic shear experiments. Accordingly, we consider the case where $\ell$-modes up to $\ell_{\rm max} = 5000$ are included in the survey, as this is a requirement for such an experiment to achieve its precision goals with weak lensing.

\begin{table}[t]
    \centering
    \caption{Photometric redshift probability distribution parameters used in this work, together with their values. The functional form of the distribution is stated in equation (\ref{eq:pphot}). These are chosen to be consistent with \cite{ISTFpap}.}
    \begin{tabular}{c c}
    \hline\hline
    Parameter & Value\\
    \hline
    $c_{\rm b}$ & 1.0\\
    $z_{\rm b}$ & 0.0\\
    $\sigma_{\rm b}$ & 0.05\\
    $c_{\rm o}$ & 1.0\\
    $z_{\rm o}$ & 0.1\\
    $\sigma_{\rm o}$ & 0.05\\
    $f_{\rm out}$ & 0.1\\
    \hline\hline
    \end{tabular}
    \label{tab:phphotparams}
\end{table}

A \emph{Euclid}-like survey will observe an area such that $f_{\rm sky} = 0.36$, and have a galaxy surface density of $\bar{n}_{\rm g}=30$ arcmin$^{-2}$. We take the intrinsic variance of unlensed galaxy ellipticities to consist of two components. These individual components are considered to both have a value of 0.21, leading to a root-mean-square intrinsic ellipticity of $\sigma_\epsilon = \sqrt{2}\times0.21 \approx 0.3$. We consider the survey to observe data in ten equi-populated redshift bins, with the following redshift bounds: \{0.001, 0.418, 0.560, 0.678, 0.789, 0.900, 1.019, 1.155, 1.324, 1.576, 2.50\}.

Taking into account photometric redshift uncertainties, the galaxy distributions for these tomographic bins are represented by the following expression:
\begin{equation}
    \label{eq:ncfht}
    {\mathcal N}_i(z) = \frac{\int_{z_i^-}^{z_i^+}{\rm d}z_{\rm p}\,\mathfrak{n}(z)p_{\rm ph}(z_{\rm p}|z)}{\int_{z_{\rm min}}^{z_{\rm max}}{\rm d}z\int_{z_i^-}^{z_i^+}{\rm d}z_{\rm p}\,\mathfrak{n}(z)p_{\rm ph}(z_{\rm p}|z)},
\end{equation}
where $z_{\rm p}$ is the photometric redshift measured, $z_i^-$ and $z_i^+$ are bounds of the $i$-th redshift bin, $z_{\rm min}$ and $z_{\rm max}$ are the minimum and maximum redshifts observed by the survey, and $\mathfrak{n}(z)$ is the underlying, true distribution of galaxies with redshift, $z$, which we model using \cite{EuclidRB}:
\begin{equation}
    \label{eq:ntrue}
    \mathfrak{n}(z) \propto \bigg(\frac{z}{z_0}\bigg)^2\,{\rm exp}\bigg[-\bigg(\frac{z}{z_0}\bigg)^{3/2}\bigg],
\end{equation}
with $z_0=z_{\rm m}/\sqrt{2}$, where $z_{\rm m}=0.9$ is the survey's median redshift, and the function $p_{\rm ph}(z_{\rm p}|z)$ encodes the probability that a galaxy with true redshift $z$ is measured instead to be at $z_{\rm p}$. Explicitly, this takes the form:
\begin{align}
\label{eq:pphot}
        p_{\rm ph}(z_{\rm p}|z) &= \frac{1-f_{\rm out}}{\sqrt{2\pi}\sigma_{\rm b}(1+z)}\,{\rm exp}\Bigg\{-\frac{1}{2}\bigg[\frac{z-c_{\rm b}z_{\rm p}-z_{\rm b}}{\sigma_{\rm b}(1+z)}\bigg]^2\Bigg\} \nonumber\\
        &+ \frac{f_{\rm out}}{\sqrt{2\pi}\sigma_{\rm o}(1+z)}\nonumber\\
        &\times{\rm exp}\Bigg\{-\frac{1}{2}\bigg[\frac{z-c_{\rm o}z_{\rm p}-z_{\rm o}}{\sigma_{\rm o}(1+z)}\bigg]^2\Bigg\},
\end{align}
where the first term on the right-hand side accounts for multiplicative and additive bias in the determination of redshifts for the fraction of sources with a well measured redshift, whereas the second term describes the effect of a fraction of catastrophic outliers, $f_{\rm out}$. Table \ref{tab:phphotparams} contains our choice of values for the parameters of this model. As a function of comoving distance, the galaxy distribution is then $n_i(\chi) = {\mathcal N}_i(z){\rm d}z/{\rm d}\chi$.

For the purposes of this investigation, we adopt a flat $w_0w_a$CDM cosmology. This choice of fiducial cosmology includes a time-varying dark energy equation-of-state. In addition, it is constituted of the following cosmological parameters: the present-day matter density parameter $\Omega_{\rm m}$, the present-day density of baryonic matter $\Omega_{\rm b}$, the amplitude of density fluctuations on 8 $h^{-1}$Mpc scales $\sigma_8$, the spectral index $n_{\rm s}$, the Hubble parameter $h=H_0/100$km\:s$^{-1}$Mpc$^{-1}$, the present-day dark energy equation of state value $w_0$, and the high-redshift value of the dark energy equation of state $w_a$. Additionally, we include massive neutrinos such that the sum of their masses $\sum m_\nu\ne 0$. Our choice of fiducial values for these parameters is given in Table \ref{tab:cosmology}. We obtain the power spectrum of the matter density contrast using the publicly available \texttt{CAMB}\footnote{\url{https://camb.info/}} code \cite{cambpap}. The non-linear part of the matter power spectrum is obtained using \texttt{Halofit} \cite{Takahashi12} and by including the additional corrections of \cite{MeadHF}. In order to calculate comoving distances, we additionally make use of \texttt{Astropy}\footnote{\url{http://www.astropy.org}} \cite{astropy1, astropy2}. The matter bispectrum required by equation (\ref{eq:Bdopp}) is computed via the \texttt{BiHalofit} model \cite{bihalofit} and package\footnote{\url{http://cosmo.phys.hirosaki-u.ac.jp/takahasi/codes_e.htm}}. In order to calculate the IA contributions at both the two--point and three--point levels, we use the following values for the NLA model: $\mathcal{A}_{\rm IA}=1.72$ and $\mathcal{C}_{\rm IA}=0.0134$ \cite{ISTFpap}. The Fisher matrix constructed in our analysis includes $\Omega_{\rm m}, \Omega_{\rm b}, h, n_{\rm s}, \sigma_8, w_0, w_a,$ and $\mathcal{A}_{\rm IA}$.

\begin{table}[t]
\centering
\caption{Choice of fiducial $w_0w_a$CDM cosmological parameter values used for this investigation. These values were chosen for consistency with \citep{ISTFpap}.}
\label{tab:cosmology}
\begin{tabular}{c c}
\hline\hline
Cosmological Parameter & Fiducial Value\\
\hline
$\Omega_{\rm m}$ & 0.32 \\
$\Omega_{\rm b}$ & 0.05 \\
$h$ & 0.67 \\
$n_{\rm s}$ & 0.96 \\
$\sigma_8$ & 0.816 \\
$\sum m_\nu$ (eV) & 0.06 \\
$w_0$ & $-1$ \\
$w_a$ & 0  \\
\hline\hline
\end{tabular}
\end{table}

\section{\label{sec:results}Results and Discussion}

\begin{figure*}[ht]
    \centering
    \includegraphics[width=1.0\linewidth]{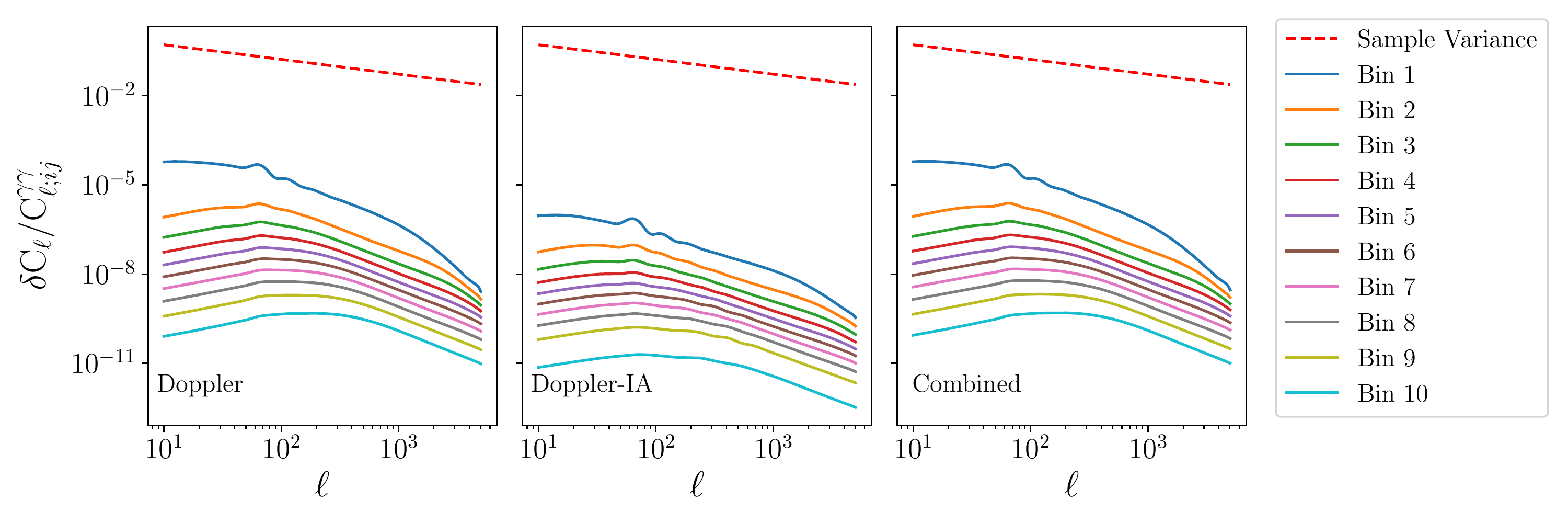}
    \caption{Relative magnitude of Doppler-shift corrections as a proportion of shear angular power spectra, for the auto-correlations of 10 equipopulated tomographic redshift bins for a \emph{Euclid}-like survey. The bin edges are: \{0.001, 0.418, 0.560, 0.678, 0.789, 0.900, 1.019, 1.155, 1.324, 1.576, 2.50\}. The left--most panel shows the Doppler--shear correction, while the central panel shows the Doppler--IA term, and the right--most panel displays the combined corrections. The sample variance is also shown, for comparison. As would be expected, the effect of Doppler-shift is greatest at low redshift, and decreases as redshift increases. In all cases, both corrections are several orders-of-magnitude below sample variance, suggesting these terms are unlikely to be significant for Stage IV surveys.}
    \label{fig:dcls_dopp}
\end{figure*}

\begin{table}[t]
\centering
\caption{Predicted $1\sigma$ cosmological parameter constraints obtained from cosmic shear power spectra for a \emph{Euclid}-like survey, together with the biases in the inferred parameter values resulting from neglecting the Doppler-shift correction. Constraints and biases are obtained using the Fisher matrix formalism, and the choice of fiducial cosmology is stated in Table \ref{tab:cosmology}. Biases are only considered significant if they exceed $0.25\times1\sigma$, as at this point the $2\sigma$ parameter constraints would overlap by less than 90$\%$. All biases reported here are well below that threshold, suggesting these corrections can be safely neglected for Stage IV experiments.}
\label{tab:biases}
\begin{tabular}{c c c c c}
\hline\hline
Cosmological & Uncertainty & Doppler & Doppler-IA\\
Parameter & (1$\sigma$) & Bias/$1\sigma$ & Bias/$1\sigma$\\
\hline
$\Omega_{\rm m}$ & 0.0089 & $1.4\times10^{-3}$ & $3.4\times10^{-6}$\\
$\Omega_{\rm b}$ & 0.020 & $-1.8\times10^{-4}$ & $-4.0\times10^{-7}$\\
$h$ & 0.12 & $-2.8\times10^{-4}$ & $-1.2\times10^{-6}$\\
$n_{\rm s}$ & 0.028 & $-1.3\times10^{-4}$ & $5.2\times10^{-7}$\\
$\sigma_8$ & 0.0094 & $-1.1\times10^{-3}$ & $-2.6\times10^{-6}$\\
$w_0$ & 0.11 & $1.2\times10^{-3}$ & $2.9\times10^{-6}$\\
$w_a$ & 0.32 & $-5.6\times10^{-4}$ & $-1.9\times10^{-6}$\\
\hline\hline
\end{tabular}
\end{table}

Within this section, we present the the effect of neglecting Doppler-shift on the cosmology that will be carried out with a \emph{Euclid}-like survey. Firstly, we show the magnitude of the Doppler and Doppler-IA corrections relative to the predicted the cosmic shear power spectra for such a survey. We then report the resulting biases on the inferred cosmological parameters that would result from ignoring these corrections.

In Figure \ref{fig:dcls_dopp}, we show the magnitude of the Doppler and Doppler-IA correction terms, relative to the cosmic shear angular power spectra, for the auto-correlation spectra of all tomographic bins for a \emph{Euclid}-like survey. Here, the two corrections are shown both separately and when combined. Additionally, the sample variance, given by \cite{SVref} $\Delta C_\ell/C_\ell = \sqrt{2}\left[f_{\mathrm{sky}}(2 \ell+1)\right]^{-1 / 2}$, is also shown for reference. From this graph, we see that the impact of Doppler-shift decreases as the redshift range of the tomographic bin probed increases. This is a consequence of the accelerating expansion of the Universe \cite{Planck18}, as accordingly we expect the relative Doppler-shift to be greater at lower redshifts. However, across the entire redshift and $\ell$ range of the survey, we observe that both correction terms remain several orders-of-magnitude below sample variance; consistent with the findings of \cite{CLSch18}. This suggests that these corrections may be able to be safely neglected for upcoming surveys.

To provide more in--depth insight into whether these terms can be neglected for Stage IV surveys, Table \ref{tab:biases} shows the biases that would result in the inferred cosmological parameter values, if the Doppler-shift effects were to be neglected. Also shown here are the predicted parameter constraints for a \emph{Euclid}-like survey. Both the predicted constraints and biases were calculated using the Fisher formalism described in Section \ref{subsec:fisher}. From this table, we see that all of the resulting biases are at sub-percent level. Given that a bias must exceed $0.25\sigma$, in order to typically be considered significant -- as at this point the biased and unbiased parameter constraints would overlap by less than $90\%$ -- we can safely conclude that the effect of Doppler-shift on the cosmic shear angular power spectrum can be neglected for Stage IV experiments.

\section{\label{sec:conclusions}Conclusions}

Within this paper, we have explored the impact of Doppler-shift on the cosmic shear angular power spectra that will obtained from Stage IV surveys. Adopting modelling specifics for a \emph{Euclid}-like survey, we calculated the three--point corrections to the shear angular power spectra that result from the perturbation to the observed shear from Doppler-shift. Additionally, we demonstrated how this perturbation interacts with IA terms, and calculated the resulting Doppler-IA correction for the shear angular power spectrum. Both of these additional corrections were shown to be several orders-of-magnitude smaller than sample variance, suggesting these corrections could be safely neglected.

In order to explicitly check whether these corrections resulted in any significant biases at the cosmological parameter level, we propagated these through a Fisher matrix calculation. We found that all resulting biases were of the sub--percent level, confirming that, in isolation, Doppler-shift does not need to be taken into account for cosmic shear analyses in Stage IV weak lensing surveys. 

However, we note that it is possible that when combined with multiple other neglected approximations, the total magnitude of the corrections may result in significant biases. A comprehensive investigation of \emph{all} weak lensing approximations is necessary to test this. Additionally, while this effect does not significantly affect the cosmic shear power spectrum, it can be detected in other forms in Stage IV surveys. If the convergence is directly probed, a significant contribution to the observed convergence signal from this Doppler-shift can be detected \cite{BaconDopp}. Furthermore, this Doppler-shift of source redshifts could also result in detectable contributions in cross-correlations with other probes that depend on the peculiar velocity of overdensities, for example the Kinematic Sunyaev–Zeldovich effect (see e.g. \cite{kSZ1, kSZ2}).

\begin{acknowledgments}

ACD acknowledges the support of the Royal Society. TDK acknowledges funding from the European Union’s Horizon 2020 research and innovation programme under grant agreement No. 776247.
\end{acknowledgments}

\appendix*
\section{\label{ap:limber} Extended Limber Approximation for Doppler Correction}

While the extended Limber approximation \cite{ExtendedLimber} can be readily applied to the $\widetilde{\gamma}^{\rm like}_{i; \ell m}$ Doppler term of equation (\ref{eq:gdoppsph}), the $\widetilde{\kappa}^{\rm like}_{i; \ell m}$ term from equation (\ref{eq:kdoppsph}) presents complications. This is due to the additional factor of $k$, and the presence of the derivative of a spherical Bessel function.

In order to apply the Limber approximation for this case, we can begin by recognizing that:
\begin{align}
    \label{eq:j_deriv}
    j'_\ell(k\chi) = \frac{\ell}{k\chi} j_\ell(k\chi) - j_{\ell + 1}(k\chi).
\end{align}
Now, we can see how we obtain the weight described in equation (\ref{eq:Wkdopp}) by following the derivation of LoVerde and Afshordi \cite{ExtendedLimber} (referred to as LA from here--on) for an angular power spectrum where one of the fields probed is $\widetilde{\kappa}^{\rm like}_{i; \ell m}$. Here we only detail the two-point case for simplicity and brevity, however it is straightforward to generalise this to the three-point case; particularly given that a bispectrum can typically be expressed as a linear combination of power spectra \cite{Frypap, ScocCouch, Gilmarin, bihalofit}. Equation (5) of LA, in our case, would read:
\begin{align}
    \label{eq:e5}
    C_{A\kappa^{\rm like}} &= \int{\rm d}k P_{A\delta} \int{\rm d}\chi_1 \frac{F_A}{\sqrt{\chi_1}} J_{\ell + 1/2} (k\chi_1) \nonumber \\
    &\int{\rm d}\chi_2 \frac{F_{\kappa^{\rm like}}}{\sqrt{\chi_2}} \nonumber\\
    &\times \bigg[\frac{\ell}{k\chi}J_{\ell + 1/2} (k\chi_2)- J_{\ell + 3/2} (k\chi_2)\bigg],
\end{align}
where $J_{\ell}$ is the Bessel function of the $\ell$-th order, $F_A$ is the projection kernel for field $A$, and:
\begin{align}
    \label{eq:Fk}
    F_{\kappa^{\rm like}}(\chi) = \frac{c}{\chi^2 H(\chi) a(\chi)} n(\chi).
\end{align}
Now, following the procedure of LA through to equation (13) of that work and retaining only terms to the first-order, we obtain:
\begin{align}
    \label{eq:Climbapp}
    C_{A\kappa^{\rm like}} &= \int \frac{{\rm d}\chi}{\chi^2} \frac{\chi}{(\ell + 1/2)}\left[\frac{\ell}{(\ell + 1/2)} - \frac{(\ell + 1/2)}{(\ell + 3/2)}\right] \nonumber\\
    &\times F_{\kappa^{\rm like}}(\chi) F_A(\chi) P_{A\delta}\left(\frac{(\ell + 1/2)}{\chi}\right) \nonumber \\
    &= \int \frac{{\rm d}\chi}{\chi^2} W^{\kappa\nu}_i(\chi, \ell) F_A(\chi) \nonumber\\
    &\times P_{A\delta}\left(\frac{(\ell + 1/2)}{\chi}\right).
\end{align}
When performed at the three-point level, this calculation gives the Limber approximated bispectra of equation (\ref{eq:Bdopp}) and equation (\ref{eq:B-D-IA}).

\bibliography{apssamp}

\providecommand{\noopsort}[1]{}\providecommand{\singleletter}[1]{#1}%
\begin{thebibliography}{35}%
\makeatletter
\providecommand \@ifxundefined [1]{%
 \@ifx{#1\undefined}
}%
\providecommand \@ifnum [1]{%
 \ifnum #1\expandafter \@firstoftwo
 \else \expandafter \@secondoftwo
 \fi
}%
\providecommand \@ifx [1]{%
 \ifx #1\expandafter \@firstoftwo
 \else \expandafter \@secondoftwo
 \fi
}%
\providecommand \natexlab [1]{#1}%
\providecommand \enquote  [1]{``#1''}%
\providecommand \bibnamefont  [1]{#1}%
\providecommand \bibfnamefont [1]{#1}%
\providecommand \citenamefont [1]{#1}%
\providecommand \href@noop [0]{\@secondoftwo}%
\providecommand \href [0]{\begingroup \@sanitize@url \@href}%
\providecommand \@href[1]{\@@startlink{#1}\@@href}%
\providecommand \@@href[1]{\endgroup#1\@@endlink}%
\providecommand \@sanitize@url [0]{\catcode `\\12\catcode `\$12\catcode
  `\&12\catcode `\#12\catcode `\^12\catcode `\_12\catcode `\%12\relax}%
\providecommand \@@startlink[1]{}%
\providecommand \@@endlink[0]{}%
\providecommand \url  [0]{\begingroup\@sanitize@url \@url }%
\providecommand \@url [1]{\endgroup\@href {#1}{\urlprefix }}%
\providecommand \urlprefix  [0]{URL }%
\providecommand \Eprint [0]{\href }%
\providecommand \doibase [0]{http://dx.doi.org/}%
\providecommand \selectlanguage [0]{\@gobble}%
\providecommand \bibinfo  [0]{\@secondoftwo}%
\providecommand \bibfield  [0]{\@secondoftwo}%
\providecommand \translation [1]{[#1]}%
\providecommand \BibitemOpen [0]{}%
\providecommand \bibitemStop [0]{}%
\providecommand \bibitemNoStop [0]{.\EOS\space}%
\providecommand \EOS [0]{\spacefactor3000\relax}%
\providecommand \BibitemShut  [1]{\csname bibitem#1\endcsname}%
\let\auto@bib@innerbib\@empty
\bibitem [{\citenamefont {{Albrecht}}\ \emph {et~al.}(2006)\citenamefont
  {{Albrecht}}, \citenamefont {{Bernstein}}, \citenamefont {{Cahn}},
  \citenamefont {{Freedman}}, \citenamefont {{Hewitt}}, \citenamefont {{Hu}},
  \citenamefont {{Huth}}, \citenamefont {{Kamionkowski}}, \citenamefont
  {{Kolb}}, \citenamefont {{Knox}} \emph {et~al.}}]{DETFrep}%
  \BibitemOpen
  \bibfield  {author} {\bibinfo {author} {\bibfnamefont {A.}~\bibnamefont
  {{Albrecht}}}, \bibinfo {author} {\bibfnamefont {G.}~\bibnamefont
  {{Bernstein}}}, \bibinfo {author} {\bibfnamefont {R.}~\bibnamefont {{Cahn}}},
  \bibinfo {author} {\bibfnamefont {W.~L.}\ \bibnamefont {{Freedman}}},
  \bibinfo {author} {\bibfnamefont {J.}~\bibnamefont {{Hewitt}}}, \bibinfo
  {author} {\bibfnamefont {W.}~\bibnamefont {{Hu}}}, \bibinfo {author}
  {\bibfnamefont {J.}~\bibnamefont {{Huth}}}, \bibinfo {author} {\bibfnamefont
  {M.}~\bibnamefont {{Kamionkowski}}}, \bibinfo {author} {\bibfnamefont
  {E.~W.}\ \bibnamefont {{Kolb}}}, \bibinfo {author} {\bibfnamefont
  {L.}~\bibnamefont {{Knox}}},  \emph {et~al.},\ }\href@noop {} {\bibfield
  {journal} {\bibinfo  {journal} {arXiv e-prints}\ ,\ \bibinfo {eid}
  {astro-ph/0609591}} (\bibinfo {year} {2006})},\ \Eprint
  {http://arxiv.org/abs/astro-ph/0609591} {arXiv:astro-ph/0609591
  [astro-ph.CO]} \BibitemShut {NoStop}%
\bibitem [{\citenamefont {{Heymans}}\ \emph {et~al.}(2012)\citenamefont
  {{Heymans}}, \citenamefont {{Van Waerbeke}}, \citenamefont {{Miller}},
  \citenamefont {{Erben}}, \citenamefont {{Hildebrandt}}, \citenamefont
  {{Hoekstra}}, \citenamefont {{Kitching}}, \citenamefont {{Mellier}},
  \citenamefont {{Simon}}, \citenamefont {{Bonnett}} \emph
  {et~al.}}]{cfhtmain}%
  \BibitemOpen
  \bibfield  {author} {\bibinfo {author} {\bibfnamefont {C.}~\bibnamefont
  {{Heymans}}}, \bibinfo {author} {\bibfnamefont {L.}~\bibnamefont {{Van
  Waerbeke}}}, \bibinfo {author} {\bibfnamefont {L.}~\bibnamefont {{Miller}}},
  \bibinfo {author} {\bibfnamefont {T.}~\bibnamefont {{Erben}}}, \bibinfo
  {author} {\bibfnamefont {H.}~\bibnamefont {{Hildebrandt}}}, \bibinfo {author}
  {\bibfnamefont {H.}~\bibnamefont {{Hoekstra}}}, \bibinfo {author}
  {\bibfnamefont {T.~D.}\ \bibnamefont {{Kitching}}}, \bibinfo {author}
  {\bibfnamefont {Y.}~\bibnamefont {{Mellier}}}, \bibinfo {author}
  {\bibfnamefont {P.}~\bibnamefont {{Simon}}}, \bibinfo {author} {\bibfnamefont
  {C.}~\bibnamefont {{Bonnett}}},  \emph {et~al.},\ }\href {\doibase
  10.1111/j.1365-2966.2012.21952.x} {\bibfield  {journal} {\bibinfo  {journal}
  {Mon. Not. R. Astron. Soc}\ }\textbf {\bibinfo {volume} {427}},\ \bibinfo
  {pages} {146} (\bibinfo {year} {2012})}\BibitemShut {NoStop}%
\bibitem [{\citenamefont {{Dark Energy Survey Collaboration}}(2005)}]{DESpap}%
  \BibitemOpen
  \bibfield  {author} {\bibinfo {author} {\bibnamefont {{Dark Energy Survey
  Collaboration}}},\ }\href@noop {} {\bibfield  {journal} {\bibinfo  {journal}
  {arXiv e-prints}\ ,\ \bibinfo {eid} {astro-ph/0510346}} (\bibinfo {year}
  {2005})},\ \Eprint {http://arxiv.org/abs/astro-ph/0510346}
  {arXiv:astro-ph/0510346 [astro-ph.CO]} \BibitemShut {NoStop}%
\bibitem [{\citenamefont {{Giblin}}\ \emph {et~al.}(2020)\citenamefont
  {{Giblin}}, \citenamefont {{Heymans}}, \citenamefont {{Asgari}},
  \citenamefont {{Hildebrandt}}, \citenamefont {{Hoekstra}}, \citenamefont
  {{Joachimi}}, \citenamefont {{Kannawadi}}, \citenamefont {{Kuijken}},
  \citenamefont {{Lin}}, \citenamefont {{Miller}} \emph {et~al.}}]{kids1000}%
  \BibitemOpen
  \bibfield  {author} {\bibinfo {author} {\bibfnamefont {B.}~\bibnamefont
  {{Giblin}}}, \bibinfo {author} {\bibfnamefont {C.}~\bibnamefont {{Heymans}}},
  \bibinfo {author} {\bibfnamefont {M.}~\bibnamefont {{Asgari}}}, \bibinfo
  {author} {\bibfnamefont {H.}~\bibnamefont {{Hildebrandt}}}, \bibinfo {author}
  {\bibfnamefont {H.}~\bibnamefont {{Hoekstra}}}, \bibinfo {author}
  {\bibfnamefont {B.}~\bibnamefont {{Joachimi}}}, \bibinfo {author}
  {\bibfnamefont {A.}~\bibnamefont {{Kannawadi}}}, \bibinfo {author}
  {\bibfnamefont {K.}~\bibnamefont {{Kuijken}}}, \bibinfo {author}
  {\bibfnamefont {C.-A.}\ \bibnamefont {{Lin}}}, \bibinfo {author}
  {\bibfnamefont {L.}~\bibnamefont {{Miller}}},  \emph {et~al.},\ }\href@noop
  {} {\bibfield  {journal} {\bibinfo  {journal} {arXiv e-prints}\ ,\ \bibinfo
  {eid} {arXiv:2007.01845}} (\bibinfo {year} {2020})},\ \Eprint
  {http://arxiv.org/abs/2007.01845} {arXiv:2007.01845 [astro-ph.CO]}
  \BibitemShut {NoStop}%
\bibitem [{\citenamefont {{Planck Collaboration}}\ \emph
  {et~al.}(2018)\citenamefont {{Planck Collaboration}}, \citenamefont
  {{Aghanim}}, \citenamefont {{Akrami}}, \citenamefont {{Ashdown}},
  \citenamefont {{Aumont}}, \citenamefont {{Baccigalupi}}, \citenamefont
  {{Ballardini}}, \citenamefont {{Banday}}, \citenamefont {{Barreiro}},
  \citenamefont {{Bartolo}} \emph {et~al.}}]{Planck18}%
  \BibitemOpen
  \bibfield  {author} {\bibinfo {author} {\bibnamefont {{Planck
  Collaboration}}}, \bibinfo {author} {\bibfnamefont {N.}~\bibnamefont
  {{Aghanim}}}, \bibinfo {author} {\bibfnamefont {Y.}~\bibnamefont {{Akrami}}},
  \bibinfo {author} {\bibfnamefont {M.}~\bibnamefont {{Ashdown}}}, \bibinfo
  {author} {\bibfnamefont {J.}~\bibnamefont {{Aumont}}}, \bibinfo {author}
  {\bibfnamefont {C.}~\bibnamefont {{Baccigalupi}}}, \bibinfo {author}
  {\bibfnamefont {M.}~\bibnamefont {{Ballardini}}}, \bibinfo {author}
  {\bibfnamefont {A.~J.}\ \bibnamefont {{Banday}}}, \bibinfo {author}
  {\bibfnamefont {R.~B.}\ \bibnamefont {{Barreiro}}}, \bibinfo {author}
  {\bibfnamefont {N.}~\bibnamefont {{Bartolo}}},  \emph {et~al.},\ }\href@noop
  {} {\bibfield  {journal} {\bibinfo  {journal} {arXiv e-prints}\ ,\ \bibinfo
  {eid} {arXiv:1807.06209}} (\bibinfo {year} {2018})},\ \Eprint
  {http://arxiv.org/abs/1807.06209} {arXiv:1807.06209 [astro-ph.CO]}
  \BibitemShut {NoStop}%
\bibitem [{\citenamefont {{Laureijs}}\ \emph {et~al.}(2011)\citenamefont
  {{Laureijs}}, \citenamefont {{Amiaux}}, \citenamefont {{Arduini}},
  \citenamefont {{Augu{\`e}res}}, \citenamefont {{Brinchmann}}, \citenamefont
  {{Cole}}, \citenamefont {{Cropper}}, \citenamefont {{Dabin}}, \citenamefont
  {{Duvet}}, \citenamefont {{Ealet}} \emph {et~al.}}]{EuclidRB}%
  \BibitemOpen
  \bibfield  {author} {\bibinfo {author} {\bibfnamefont {R.}~\bibnamefont
  {{Laureijs}}}, \bibinfo {author} {\bibfnamefont {J.}~\bibnamefont
  {{Amiaux}}}, \bibinfo {author} {\bibfnamefont {S.}~\bibnamefont {{Arduini}}},
  \bibinfo {author} {\bibfnamefont {J.-L.}\ \bibnamefont {{Augu{\`e}res}}},
  \bibinfo {author} {\bibfnamefont {J.}~\bibnamefont {{Brinchmann}}}, \bibinfo
  {author} {\bibfnamefont {R.}~\bibnamefont {{Cole}}}, \bibinfo {author}
  {\bibfnamefont {M.}~\bibnamefont {{Cropper}}}, \bibinfo {author}
  {\bibfnamefont {C.}~\bibnamefont {{Dabin}}}, \bibinfo {author} {\bibfnamefont
  {L.}~\bibnamefont {{Duvet}}}, \bibinfo {author} {\bibfnamefont
  {A.}~\bibnamefont {{Ealet}}},  \emph {et~al.},\ }\href@noop {} {\bibfield
  {journal} {\bibinfo  {journal} {arXiv e-prints}\ ,\ \bibinfo {eid}
  {arXiv:1110.3193}} (\bibinfo {year} {2011})},\ \Eprint
  {http://arxiv.org/abs/1110.3193} {arXiv:1110.3193 [astro-ph.CO]} \BibitemShut
  {NoStop}%
\bibitem [{\citenamefont {{Akeson}}\ \emph {et~al.}(2019)\citenamefont
  {{Akeson}}, \citenamefont {{Armus}}, \citenamefont {{Bachelet}},
  \citenamefont {{Bailey}}, \citenamefont {{Bartusek}}, \citenamefont
  {{Bellini}}, \citenamefont {{Benford}}, \citenamefont {{Bennett}},
  \citenamefont {{Bhattacharya}}, \citenamefont {{Bohlin}} \emph
  {et~al.}}]{WFIRSTpap}%
  \BibitemOpen
  \bibfield  {author} {\bibinfo {author} {\bibfnamefont {R.}~\bibnamefont
  {{Akeson}}}, \bibinfo {author} {\bibfnamefont {L.}~\bibnamefont {{Armus}}},
  \bibinfo {author} {\bibfnamefont {E.}~\bibnamefont {{Bachelet}}}, \bibinfo
  {author} {\bibfnamefont {V.}~\bibnamefont {{Bailey}}}, \bibinfo {author}
  {\bibfnamefont {L.}~\bibnamefont {{Bartusek}}}, \bibinfo {author}
  {\bibfnamefont {A.}~\bibnamefont {{Bellini}}}, \bibinfo {author}
  {\bibfnamefont {D.}~\bibnamefont {{Benford}}}, \bibinfo {author}
  {\bibfnamefont {D.}~\bibnamefont {{Bennett}}}, \bibinfo {author}
  {\bibfnamefont {A.}~\bibnamefont {{Bhattacharya}}}, \bibinfo {author}
  {\bibfnamefont {R.}~\bibnamefont {{Bohlin}}},  \emph {et~al.},\ }\href@noop
  {} {\bibfield  {journal} {\bibinfo  {journal} {arXiv e-prints}\ ,\ \bibinfo
  {eid} {arXiv:1902.05569}} (\bibinfo {year} {2019})},\ \Eprint
  {http://arxiv.org/abs/1902.05569} {arXiv:1902.05569 [astro-ph.IM]}
  \BibitemShut {NoStop}%
\bibitem [{\citenamefont {{LSST Science Collaboration}}\ \emph
  {et~al.}(2009)\citenamefont {{LSST Science Collaboration}}, \citenamefont
  {{Abell}}, \citenamefont {{Allison}}, \citenamefont {{Anderson}},
  \citenamefont {{Andrew}}, \citenamefont {{Angel}}, \citenamefont {{Armus}},
  \citenamefont {{Arnett}}, \citenamefont {{Asztalos}}, \citenamefont
  {{Axelrod}} \emph {et~al.}}]{LSSTpap}%
  \BibitemOpen
  \bibfield  {author} {\bibinfo {author} {\bibnamefont {{LSST Science
  Collaboration}}}, \bibinfo {author} {\bibfnamefont {P.~A.}\ \bibnamefont
  {{Abell}}}, \bibinfo {author} {\bibfnamefont {J.}~\bibnamefont {{Allison}}},
  \bibinfo {author} {\bibfnamefont {S.~F.}\ \bibnamefont {{Anderson}}},
  \bibinfo {author} {\bibfnamefont {J.~R.}\ \bibnamefont {{Andrew}}}, \bibinfo
  {author} {\bibfnamefont {J.~R.~P.}\ \bibnamefont {{Angel}}}, \bibinfo
  {author} {\bibfnamefont {L.}~\bibnamefont {{Armus}}}, \bibinfo {author}
  {\bibfnamefont {D.}~\bibnamefont {{Arnett}}}, \bibinfo {author}
  {\bibfnamefont {S.~J.}\ \bibnamefont {{Asztalos}}}, \bibinfo {author}
  {\bibfnamefont {T.~S.}\ \bibnamefont {{Axelrod}}},  \emph {et~al.},\
  }\href@noop {} {\bibfield  {journal} {\bibinfo  {journal} {arXiv e-prints}\
  ,\ \bibinfo {eid} {arXiv:0912.0201}} (\bibinfo {year} {2009})},\ \Eprint
  {http://arxiv.org/abs/0912.0201} {arXiv:0912.0201 [astro-ph.IM]} \BibitemShut
  {NoStop}%
\bibitem [{\citenamefont {{Bernardeau}}\ \emph {et~al.}(2010)\citenamefont
  {{Bernardeau}}, \citenamefont {{Bonvin}},\ and\ \citenamefont
  {{Vernizzi}}}]{DopplerBernardeau}%
  \BibitemOpen
  \bibfield  {author} {\bibinfo {author} {\bibfnamefont {F.}~\bibnamefont
  {{Bernardeau}}}, \bibinfo {author} {\bibfnamefont {C.}~\bibnamefont
  {{Bonvin}}}, \ and\ \bibinfo {author} {\bibfnamefont {F.}~\bibnamefont
  {{Vernizzi}}},\ }\href {\doibase 10.1103/PhysRevD.81.083002} {\bibfield
  {journal} {\bibinfo  {journal} {Phys. Rev. D}\ }\textbf {\bibinfo {volume}
  {81}},\ \bibinfo {eid} {083002} (\bibinfo {year} {2010})}\BibitemShut
  {NoStop}%
\bibitem [{\citenamefont {{Cuesta-Lazaro}}\ \emph {et~al.}(2018)\citenamefont
  {{Cuesta-Lazaro}}, \citenamefont {{Quera-Bofarull}}, \citenamefont
  {{Reischke}},\ and\ \citenamefont {{Sch{\"a}fer}}}]{CLSch18}%
  \BibitemOpen
  \bibfield  {author} {\bibinfo {author} {\bibfnamefont {C.}~\bibnamefont
  {{Cuesta-Lazaro}}}, \bibinfo {author} {\bibfnamefont {A.}~\bibnamefont
  {{Quera-Bofarull}}}, \bibinfo {author} {\bibfnamefont {R.}~\bibnamefont
  {{Reischke}}}, \ and\ \bibinfo {author} {\bibfnamefont {B.~M.}\ \bibnamefont
  {{Sch{\"a}fer}}},\ }\href {\doibase 10.1093/mnras/sty672} {\bibfield
  {journal} {\bibinfo  {journal} {Mon. Not. R. Astron. Soc}\ }\textbf {\bibinfo
  {volume} {477}},\ \bibinfo {pages} {741} (\bibinfo {year}
  {2018})}\BibitemShut {NoStop}%
\bibitem [{\citenamefont {{Euclid Collaboration}}\ \emph
  {et~al.}(2019)\citenamefont {{Euclid Collaboration}}, \citenamefont
  {{Blanchard}}, \citenamefont {{Camera}}, \citenamefont {{Carbone}},
  \citenamefont {{Cardone}}, \citenamefont {{Casas}}, \citenamefont
  {{Ili{\'c}}}, \citenamefont {{Kilbinger}}, \citenamefont {{Kitching}},
  \citenamefont {{Kunz}} \emph {et~al.}}]{ISTFpap}%
  \BibitemOpen
  \bibfield  {author} {\bibinfo {author} {\bibnamefont {{Euclid
  Collaboration}}}, \bibinfo {author} {\bibfnamefont {A.}~\bibnamefont
  {{Blanchard}}}, \bibinfo {author} {\bibfnamefont {S.}~\bibnamefont
  {{Camera}}}, \bibinfo {author} {\bibfnamefont {C.}~\bibnamefont {{Carbone}}},
  \bibinfo {author} {\bibfnamefont {V.~F.}\ \bibnamefont {{Cardone}}}, \bibinfo
  {author} {\bibfnamefont {S.}~\bibnamefont {{Casas}}}, \bibinfo {author}
  {\bibfnamefont {S.}~\bibnamefont {{Ili{\'c}}}}, \bibinfo {author}
  {\bibfnamefont {M.}~\bibnamefont {{Kilbinger}}}, \bibinfo {author}
  {\bibfnamefont {T.}~\bibnamefont {{Kitching}}}, \bibinfo {author}
  {\bibfnamefont {M.}~\bibnamefont {{Kunz}}},  \emph {et~al.},\ }\href@noop {}
  {\bibfield  {journal} {\bibinfo  {journal} {arXiv e-prints}\ ,\ \bibinfo
  {eid} {arXiv:1910.09273}} (\bibinfo {year} {2019})},\ \Eprint
  {http://arxiv.org/abs/1910.09273} {arXiv:1910.09273 [astro-ph.CO]}
  \BibitemShut {NoStop}%
\bibitem [{\citenamefont {{Deshpande}}\ \emph
  {et~al.}(2020{\natexlab{a}})\citenamefont {{Deshpande}}, \citenamefont
  {{Kitching}}, \citenamefont {{Cardone}}, \citenamefont {{Taylor}},
  \citenamefont {{Casas}}, \citenamefont {{Camera}}, \citenamefont {{Carbone}},
  \citenamefont {{Kilbinger}}, \citenamefont {{Pettorino}}, \citenamefont
  {{Sakr}} \emph {et~al.}}]{Deshpap}%
  \BibitemOpen
  \bibfield  {author} {\bibinfo {author} {\bibfnamefont {A.~C.}\ \bibnamefont
  {{Deshpande}}}, \bibinfo {author} {\bibfnamefont {T.~D.}\ \bibnamefont
  {{Kitching}}}, \bibinfo {author} {\bibfnamefont {V.~F.}\ \bibnamefont
  {{Cardone}}}, \bibinfo {author} {\bibfnamefont {P.~L.}\ \bibnamefont
  {{Taylor}}}, \bibinfo {author} {\bibfnamefont {S.}~\bibnamefont {{Casas}}},
  \bibinfo {author} {\bibfnamefont {S.}~\bibnamefont {{Camera}}}, \bibinfo
  {author} {\bibfnamefont {C.}~\bibnamefont {{Carbone}}}, \bibinfo {author}
  {\bibfnamefont {M.}~\bibnamefont {{Kilbinger}}}, \bibinfo {author}
  {\bibfnamefont {V.}~\bibnamefont {{Pettorino}}}, \bibinfo {author}
  {\bibfnamefont {Z.}~\bibnamefont {{Sakr}}},  \emph {et~al.},\ }\href
  {\doibase 10.1051/0004-6361/201937323} {\bibfield  {journal} {\bibinfo
  {journal} {Astron. Astrophys.}\ }\textbf {\bibinfo {volume} {636}},\ \bibinfo
  {eid} {A95} (\bibinfo {year} {2020}{\natexlab{a}})}\BibitemShut {NoStop}%
\bibitem [{\citenamefont {{Deshpande}}\ \emph
  {et~al.}(2020{\natexlab{b}})\citenamefont {{Deshpande}}, \citenamefont
  {{Taylor}},\ and\ \citenamefont {{Kitching}}}]{kcutrs}%
  \BibitemOpen
  \bibfield  {author} {\bibinfo {author} {\bibfnamefont {A.~C.}\ \bibnamefont
  {{Deshpande}}}, \bibinfo {author} {\bibfnamefont {P.~L.}\ \bibnamefont
  {{Taylor}}}, \ and\ \bibinfo {author} {\bibfnamefont {T.~D.}\ \bibnamefont
  {{Kitching}}},\ }\href {\doibase 10.1103/PhysRevD.102.083535} {\bibfield
  {journal} {\bibinfo  {journal} {Phys. Rev. D}\ }\textbf {\bibinfo {volume}
  {102}},\ \bibinfo {eid} {083535} (\bibinfo {year}
  {2020}{\natexlab{b}})}\BibitemShut {NoStop}%
\bibitem [{\citenamefont {{LoVerde}}\ and\ \citenamefont
  {{Afshordi}}(2008)}]{ExtendedLimber}%
  \BibitemOpen
  \bibfield  {author} {\bibinfo {author} {\bibfnamefont {M.}~\bibnamefont
  {{LoVerde}}}\ and\ \bibinfo {author} {\bibfnamefont {N.}~\bibnamefont
  {{Afshordi}}},\ }\href {\doibase 10.1103/PhysRevD.78.123506} {\bibfield
  {journal} {\bibinfo  {journal} {Phys. Rev. D}\ }\textbf {\bibinfo {volume}
  {78}},\ \bibinfo {eid} {123506} (\bibinfo {year} {2008})}\BibitemShut
  {NoStop}%
\bibitem [{\citenamefont {{Kitching}}\ \emph {et~al.}(2017)\citenamefont
  {{Kitching}}, \citenamefont {{Alsing}}, \citenamefont {{Heavens}},
  \citenamefont {{Jimenez}}, \citenamefont {{McEwen}},\ and\ \citenamefont
  {{Verde}}}]{limitsofshear17}%
  \BibitemOpen
  \bibfield  {author} {\bibinfo {author} {\bibfnamefont {T.~D.}\ \bibnamefont
  {{Kitching}}}, \bibinfo {author} {\bibfnamefont {J.}~\bibnamefont
  {{Alsing}}}, \bibinfo {author} {\bibfnamefont {A.~F.}\ \bibnamefont
  {{Heavens}}}, \bibinfo {author} {\bibfnamefont {R.}~\bibnamefont
  {{Jimenez}}}, \bibinfo {author} {\bibfnamefont {J.~D.}\ \bibnamefont
  {{McEwen}}}, \ and\ \bibinfo {author} {\bibfnamefont {L.}~\bibnamefont
  {{Verde}}},\ }\href {\doibase 10.1093/mnras/stx1039} {\bibfield  {journal}
  {\bibinfo  {journal} {Mon. Not. R. Astron. Soc}\ }\textbf {\bibinfo {volume}
  {469}},\ \bibinfo {pages} {2737} (\bibinfo {year} {2017})}\BibitemShut
  {NoStop}%
\bibitem [{\citenamefont {{Kilbinger}}(2015)}]{Kilbinger15}%
  \BibitemOpen
  \bibfield  {author} {\bibinfo {author} {\bibfnamefont {M.}~\bibnamefont
  {{Kilbinger}}},\ }\href {\doibase 10.1088/0034-4885/78/8/086901} {\bibfield
  {journal} {\bibinfo  {journal} {Rep. Prog. Phys.}\ }\textbf {\bibinfo
  {volume} {78}},\ \bibinfo {eid} {086901} (\bibinfo {year}
  {2015})}\BibitemShut {NoStop}%
\bibitem [{\citenamefont {{Joachimi}}\ \emph {et~al.}(2015)\citenamefont
  {{Joachimi}}, \citenamefont {{Cacciato}}, \citenamefont {{Kitching}},
  \citenamefont {{Leonard}}, \citenamefont {{Mandelbaum}}, \citenamefont
  {{Sch{\"a}fer}}, \citenamefont {{Sif{\'o}n}}, \citenamefont {{Hoekstra}},
  \citenamefont {{Kiessling}}, \citenamefont {{Kirk}},\ and\ \citenamefont
  {{Rassat}}}]{JoachimiIAs}%
  \BibitemOpen
  \bibfield  {author} {\bibinfo {author} {\bibfnamefont {B.}~\bibnamefont
  {{Joachimi}}}, \bibinfo {author} {\bibfnamefont {M.}~\bibnamefont
  {{Cacciato}}}, \bibinfo {author} {\bibfnamefont {T.~D.}\ \bibnamefont
  {{Kitching}}}, \bibinfo {author} {\bibfnamefont {A.}~\bibnamefont
  {{Leonard}}}, \bibinfo {author} {\bibfnamefont {R.}~\bibnamefont
  {{Mandelbaum}}}, \bibinfo {author} {\bibfnamefont {B.~M.}\ \bibnamefont
  {{Sch{\"a}fer}}}, \bibinfo {author} {\bibfnamefont {C.}~\bibnamefont
  {{Sif{\'o}n}}}, \bibinfo {author} {\bibfnamefont {H.}~\bibnamefont
  {{Hoekstra}}}, \bibinfo {author} {\bibfnamefont {A.}~\bibnamefont
  {{Kiessling}}}, \bibinfo {author} {\bibfnamefont {D.}~\bibnamefont {{Kirk}}},
  \ and\ \bibinfo {author} {\bibfnamefont {A.}~\bibnamefont {{Rassat}}},\
  }\href {\doibase 10.1007/s11214-015-0177-4} {\bibfield  {journal} {\bibinfo
  {journal} {Space Sci. Rev.}\ }\textbf {\bibinfo {volume} {193}},\ \bibinfo
  {pages} {1} (\bibinfo {year} {2015})}\BibitemShut {NoStop}%
\bibitem [{\citenamefont {{Bridle}}\ and\ \citenamefont
  {{King}}(2007)}]{NLAmodel}%
  \BibitemOpen
  \bibfield  {author} {\bibinfo {author} {\bibfnamefont {S.}~\bibnamefont
  {{Bridle}}}\ and\ \bibinfo {author} {\bibfnamefont {L.}~\bibnamefont
  {{King}}},\ }\href {\doibase 10.1088/1367-2630/9/12/444} {\bibfield
  {journal} {\bibinfo  {journal} {New J. Phys.}\ }\textbf {\bibinfo {volume}
  {9}},\ \bibinfo {pages} {444} (\bibinfo {year} {2007})}\BibitemShut {NoStop}%
\bibitem [{\citenamefont {{Bacon}}\ \emph {et~al.}(2014)\citenamefont
  {{Bacon}}, \citenamefont {{Andrianomena}}, \citenamefont {{Clarkson}},
  \citenamefont {{Bolejko}},\ and\ \citenamefont {{Maartens}}}]{BaconDopp}%
  \BibitemOpen
  \bibfield  {author} {\bibinfo {author} {\bibfnamefont {D.~J.}\ \bibnamefont
  {{Bacon}}}, \bibinfo {author} {\bibfnamefont {S.}~\bibnamefont
  {{Andrianomena}}}, \bibinfo {author} {\bibfnamefont {C.}~\bibnamefont
  {{Clarkson}}}, \bibinfo {author} {\bibfnamefont {K.}~\bibnamefont
  {{Bolejko}}}, \ and\ \bibinfo {author} {\bibfnamefont {R.}~\bibnamefont
  {{Maartens}}},\ }\href {\doibase 10.1093/mnras/stu1270} {\bibfield  {journal}
  {\bibinfo  {journal} {Mon. Not. R. Astron. Soc}\ }\textbf {\bibinfo {volume}
  {443}},\ \bibinfo {pages} {1900} (\bibinfo {year} {2014})}\BibitemShut
  {NoStop}%
\bibitem [{\citenamefont {{Scoccimarro}}\ and\ \citenamefont
  {{Couchman}}(2001)}]{ScocCouch}%
  \BibitemOpen
  \bibfield  {author} {\bibinfo {author} {\bibfnamefont {R.}~\bibnamefont
  {{Scoccimarro}}}\ and\ \bibinfo {author} {\bibfnamefont {H.~M.~P.}\
  \bibnamefont {{Couchman}}},\ }\href {\doibase
  10.1046/j.1365-8711.2001.04281.x} {\bibfield  {journal} {\bibinfo  {journal}
  {Mon. Not. R. Astron. Soc}\ }\textbf {\bibinfo {volume} {325}},\ \bibinfo
  {pages} {1312} (\bibinfo {year} {2001})}\BibitemShut {NoStop}%
\bibitem [{\citenamefont {{Tegmark}}\ \emph {et~al.}(1997)\citenamefont
  {{Tegmark}}, \citenamefont {{Taylor}},\ and\ \citenamefont
  {{Heavens}}}]{Tegmark97}%
  \BibitemOpen
  \bibfield  {author} {\bibinfo {author} {\bibfnamefont {M.}~\bibnamefont
  {{Tegmark}}}, \bibinfo {author} {\bibfnamefont {A.~N.}\ \bibnamefont
  {{Taylor}}}, \ and\ \bibinfo {author} {\bibfnamefont {A.~F.}\ \bibnamefont
  {{Heavens}}},\ }\href {\doibase 10.1086/303939} {\bibfield  {journal}
  {\bibinfo  {journal} {Astrophys. J}\ }\textbf {\bibinfo {volume} {480}},\
  \bibinfo {pages} {22} (\bibinfo {year} {1997})}\BibitemShut {NoStop}%
\bibitem [{\citenamefont {{Lin}}\ \emph {et~al.}(2019)\citenamefont {{Lin}},
  \citenamefont {{Harnois-D{\'e}raps}}, \citenamefont {{Eifler}}, \citenamefont
  {{Pospisil}}, \citenamefont {{Mandelbaum}}, \citenamefont {{Lee}},\ and\
  \citenamefont {{Singh}}}]{LSSTnongauss}%
  \BibitemOpen
  \bibfield  {author} {\bibinfo {author} {\bibfnamefont {C.-H.}\ \bibnamefont
  {{Lin}}}, \bibinfo {author} {\bibfnamefont {J.}~\bibnamefont
  {{Harnois-D{\'e}raps}}}, \bibinfo {author} {\bibfnamefont {T.}~\bibnamefont
  {{Eifler}}}, \bibinfo {author} {\bibfnamefont {T.}~\bibnamefont
  {{Pospisil}}}, \bibinfo {author} {\bibfnamefont {R.}~\bibnamefont
  {{Mandelbaum}}}, \bibinfo {author} {\bibfnamefont {A.~B.}\ \bibnamefont
  {{Lee}}}, \ and\ \bibinfo {author} {\bibfnamefont {S.}~\bibnamefont
  {{Singh}}},\ }\href@noop {} {\bibfield  {journal} {\bibinfo  {journal} {arXiv
  e-prints}\ ,\ \bibinfo {eid} {arXiv:1905.03779}} (\bibinfo {year} {2019})},\
  \Eprint {http://arxiv.org/abs/1905.03779} {arXiv:1905.03779 [astro-ph.CO]}
  \BibitemShut {NoStop}%
\bibitem [{\citenamefont {{Taylor}}\ \emph {et~al.}(2019)\citenamefont
  {{Taylor}}, \citenamefont {{Kitching}}, \citenamefont {{Alsing}},
  \citenamefont {{Wandelt}}, \citenamefont {{Feeney}},\ and\ \citenamefont
  {{McEwen}}}]{gausspeter}%
  \BibitemOpen
  \bibfield  {author} {\bibinfo {author} {\bibfnamefont {P.~L.}\ \bibnamefont
  {{Taylor}}}, \bibinfo {author} {\bibfnamefont {T.~D.}\ \bibnamefont
  {{Kitching}}}, \bibinfo {author} {\bibfnamefont {J.}~\bibnamefont
  {{Alsing}}}, \bibinfo {author} {\bibfnamefont {B.~D.}\ \bibnamefont
  {{Wandelt}}}, \bibinfo {author} {\bibfnamefont {S.~M.}\ \bibnamefont
  {{Feeney}}}, \ and\ \bibinfo {author} {\bibfnamefont {J.~D.}\ \bibnamefont
  {{McEwen}}},\ }\href {\doibase 10.1103/PhysRevD.100.023519} {\bibfield
  {journal} {\bibinfo  {journal} {Phys. Rev. D}\ }\textbf {\bibinfo {volume}
  {100}},\ \bibinfo {eid} {023519} (\bibinfo {year} {2019})}\BibitemShut
  {NoStop}%
\bibitem [{\citenamefont {{Taylor}}\ \emph {et~al.}(2007)\citenamefont
  {{Taylor}}, \citenamefont {{Kitching}}, \citenamefont {{Bacon}},\ and\
  \citenamefont {{Heavens}}}]{Fishbias}%
  \BibitemOpen
  \bibfield  {author} {\bibinfo {author} {\bibfnamefont {A.~N.}\ \bibnamefont
  {{Taylor}}}, \bibinfo {author} {\bibfnamefont {T.~D.}\ \bibnamefont
  {{Kitching}}}, \bibinfo {author} {\bibfnamefont {D.~J.}\ \bibnamefont
  {{Bacon}}}, \ and\ \bibinfo {author} {\bibfnamefont {A.~F.}\ \bibnamefont
  {{Heavens}}},\ }\href {\doibase 10.1111/j.1365-2966.2006.11257.x} {\bibfield
  {journal} {\bibinfo  {journal} {Mon. Not. R. Astron. Soc}\ }\textbf {\bibinfo
  {volume} {374}},\ \bibinfo {pages} {1377} (\bibinfo {year}
  {2007})}\BibitemShut {NoStop}%
\bibitem [{\citenamefont {{Lewis}}\ \emph {et~al.}(2000)\citenamefont
  {{Lewis}}, \citenamefont {{Challinor}},\ and\ \citenamefont
  {{Lasenby}}}]{cambpap}%
  \BibitemOpen
  \bibfield  {author} {\bibinfo {author} {\bibfnamefont {A.}~\bibnamefont
  {{Lewis}}}, \bibinfo {author} {\bibfnamefont {A.}~\bibnamefont
  {{Challinor}}}, \ and\ \bibinfo {author} {\bibfnamefont {A.}~\bibnamefont
  {{Lasenby}}},\ }\href {\doibase 10.1086/309179} {\bibfield  {journal}
  {\bibinfo  {journal} {Astrophys. J}\ }\textbf {\bibinfo {volume} {538}},\
  \bibinfo {pages} {473} (\bibinfo {year} {2000})}\BibitemShut {NoStop}%
\bibitem [{\citenamefont {{Takahashi}}\ \emph {et~al.}(2012)\citenamefont
  {{Takahashi}}, \citenamefont {{Sato}}, \citenamefont {{Nishimichi}},
  \citenamefont {{Taruya}},\ and\ \citenamefont {{Oguri}}}]{Takahashi12}%
  \BibitemOpen
  \bibfield  {author} {\bibinfo {author} {\bibfnamefont {R.}~\bibnamefont
  {{Takahashi}}}, \bibinfo {author} {\bibfnamefont {M.}~\bibnamefont {{Sato}}},
  \bibinfo {author} {\bibfnamefont {T.}~\bibnamefont {{Nishimichi}}}, \bibinfo
  {author} {\bibfnamefont {A.}~\bibnamefont {{Taruya}}}, \ and\ \bibinfo
  {author} {\bibfnamefont {M.}~\bibnamefont {{Oguri}}},\ }\href {\doibase
  10.1088/0004-637X/761/2/152} {\bibfield  {journal} {\bibinfo  {journal}
  {Astrophys. J}\ }\textbf {\bibinfo {volume} {761}},\ \bibinfo {eid} {152}
  (\bibinfo {year} {2012})}\BibitemShut {NoStop}%
\bibitem [{\citenamefont {{Mead}}\ \emph {et~al.}(2015)\citenamefont {{Mead}},
  \citenamefont {{Peacock}}, \citenamefont {{Heymans}}, \citenamefont
  {{Joudaki}},\ and\ \citenamefont {{Heavens}}}]{MeadHF}%
  \BibitemOpen
  \bibfield  {author} {\bibinfo {author} {\bibfnamefont {A.~J.}\ \bibnamefont
  {{Mead}}}, \bibinfo {author} {\bibfnamefont {J.~A.}\ \bibnamefont
  {{Peacock}}}, \bibinfo {author} {\bibfnamefont {C.}~\bibnamefont
  {{Heymans}}}, \bibinfo {author} {\bibfnamefont {S.}~\bibnamefont
  {{Joudaki}}}, \ and\ \bibinfo {author} {\bibfnamefont {A.~F.}\ \bibnamefont
  {{Heavens}}},\ }\href {\doibase 10.1093/mnras/stv2036} {\bibfield  {journal}
  {\bibinfo  {journal} {Mon. Not. R. Astron. Soc}\ }\textbf {\bibinfo {volume}
  {454}},\ \bibinfo {pages} {1958} (\bibinfo {year} {2015})}\BibitemShut
  {NoStop}%
\bibitem [{\citenamefont {{Astropy Collaboration}}\ \emph
  {et~al.}(2013)\citenamefont {{Astropy Collaboration}}, \citenamefont
  {{Robitaille}}, \citenamefont {{Tollerud}}, \citenamefont {{Greenfield}},
  \citenamefont {{Droettboom}}, \citenamefont {{Bray}}, \citenamefont
  {{Aldcroft}}, \citenamefont {{Davis}}, \citenamefont {{Ginsburg}},
  \citenamefont {{Price-Whelan}} \emph {et~al.}}]{astropy1}%
  \BibitemOpen
  \bibfield  {author} {\bibinfo {author} {\bibnamefont {{Astropy
  Collaboration}}}, \bibinfo {author} {\bibfnamefont {T.~P.}\ \bibnamefont
  {{Robitaille}}}, \bibinfo {author} {\bibfnamefont {E.~J.}\ \bibnamefont
  {{Tollerud}}}, \bibinfo {author} {\bibfnamefont {P.}~\bibnamefont
  {{Greenfield}}}, \bibinfo {author} {\bibfnamefont {M.}~\bibnamefont
  {{Droettboom}}}, \bibinfo {author} {\bibfnamefont {E.}~\bibnamefont
  {{Bray}}}, \bibinfo {author} {\bibfnamefont {T.}~\bibnamefont {{Aldcroft}}},
  \bibinfo {author} {\bibfnamefont {M.}~\bibnamefont {{Davis}}}, \bibinfo
  {author} {\bibfnamefont {A.}~\bibnamefont {{Ginsburg}}}, \bibinfo {author}
  {\bibfnamefont {A.~M.}\ \bibnamefont {{Price-Whelan}}},  \emph {et~al.},\
  }\href {\doibase 10.1051/0004-6361/201322068} {\bibfield  {journal} {\bibinfo
   {journal} {Astron. Astrophys.}\ }\textbf {\bibinfo {volume} {558}},\
  \bibinfo {eid} {A33} (\bibinfo {year} {2013})}\BibitemShut {NoStop}%
\bibitem [{\citenamefont {{Astropy Collaboration}}\ \emph
  {et~al.}(2018)\citenamefont {{Astropy Collaboration}}, \citenamefont
  {{Price-Whelan}}, \citenamefont {{Sip{\H{o}}cz}}, \citenamefont
  {{G{\"u}nther}}, \citenamefont {{Lim}}, \citenamefont {{Crawford}},
  \citenamefont {{Conseil}}, \citenamefont {{Shupe}}, \citenamefont {{Craig}},
  \citenamefont {{Dencheva}} \emph {et~al.}}]{astropy2}%
  \BibitemOpen
  \bibfield  {author} {\bibinfo {author} {\bibnamefont {{Astropy
  Collaboration}}}, \bibinfo {author} {\bibfnamefont {A.~M.}\ \bibnamefont
  {{Price-Whelan}}}, \bibinfo {author} {\bibfnamefont {B.~M.}\ \bibnamefont
  {{Sip{\H{o}}cz}}}, \bibinfo {author} {\bibfnamefont {H.~M.}\ \bibnamefont
  {{G{\"u}nther}}}, \bibinfo {author} {\bibfnamefont {P.~L.}\ \bibnamefont
  {{Lim}}}, \bibinfo {author} {\bibfnamefont {S.~M.}\ \bibnamefont
  {{Crawford}}}, \bibinfo {author} {\bibfnamefont {S.}~\bibnamefont
  {{Conseil}}}, \bibinfo {author} {\bibfnamefont {D.~L.}\ \bibnamefont
  {{Shupe}}}, \bibinfo {author} {\bibfnamefont {M.~W.}\ \bibnamefont
  {{Craig}}}, \bibinfo {author} {\bibfnamefont {N.}~\bibnamefont {{Dencheva}}},
   \emph {et~al.},\ }\href {\doibase 10.3847/1538-3881/aabc4f} {\bibfield
  {journal} {\bibinfo  {journal} {Astron. J}\ }\textbf {\bibinfo {volume}
  {156}},\ \bibinfo {eid} {123} (\bibinfo {year} {2018})}\BibitemShut {NoStop}%
\bibitem [{\citenamefont {{Takahashi}}\ \emph {et~al.}(2020)\citenamefont
  {{Takahashi}}, \citenamefont {{Nishimichi}}, \citenamefont {{Namikawa}},
  \citenamefont {{Taruya}}, \citenamefont {{Kayo}}, \citenamefont {{Osato}},
  \citenamefont {{Kobayashi}},\ and\ \citenamefont {{Shirasaki}}}]{bihalofit}%
  \BibitemOpen
  \bibfield  {author} {\bibinfo {author} {\bibfnamefont {R.}~\bibnamefont
  {{Takahashi}}}, \bibinfo {author} {\bibfnamefont {T.}~\bibnamefont
  {{Nishimichi}}}, \bibinfo {author} {\bibfnamefont {T.}~\bibnamefont
  {{Namikawa}}}, \bibinfo {author} {\bibfnamefont {A.}~\bibnamefont
  {{Taruya}}}, \bibinfo {author} {\bibfnamefont {I.}~\bibnamefont {{Kayo}}},
  \bibinfo {author} {\bibfnamefont {K.}~\bibnamefont {{Osato}}}, \bibinfo
  {author} {\bibfnamefont {Y.}~\bibnamefont {{Kobayashi}}}, \ and\ \bibinfo
  {author} {\bibfnamefont {M.}~\bibnamefont {{Shirasaki}}},\ }\href {\doibase
  10.3847/1538-4357/ab908d} {\bibfield  {journal} {\bibinfo  {journal}
  {Astrophys. J}\ }\textbf {\bibinfo {volume} {895}},\ \bibinfo {eid} {113}
  (\bibinfo {year} {2020})}\BibitemShut {NoStop}%
\bibitem [{\citenamefont {Weinberg}(2008)}]{SVref}%
  \BibitemOpen
  \bibfield  {author} {\bibinfo {author} {\bibfnamefont {S.}~\bibnamefont
  {Weinberg}},\ }\href@noop {} {\emph {\bibinfo {title} {Cosmology}}}\
  (\bibinfo  {publisher} {Oxford University Press},\ \bibinfo {year}
  {2008})\BibitemShut {NoStop}%
\bibitem [{\citenamefont {Shao}\ \emph {et~al.}(2011)\citenamefont {Shao},
  \citenamefont {Zhang}, \citenamefont {Lin}, \citenamefont {Jing},\ and\
  \citenamefont {Pan}}]{kSZ1}%
  \BibitemOpen
  \bibfield  {author} {\bibinfo {author} {\bibfnamefont {J.}~\bibnamefont
  {Shao}}, \bibinfo {author} {\bibfnamefont {P.}~\bibnamefont {Zhang}},
  \bibinfo {author} {\bibfnamefont {W.}~\bibnamefont {Lin}}, \bibinfo {author}
  {\bibfnamefont {Y.}~\bibnamefont {Jing}}, \ and\ \bibinfo {author}
  {\bibfnamefont {J.}~\bibnamefont {Pan}},\ }\href {\doibase
  10.1111/j.1365-2966.2011.18166.x} {\bibfield  {journal} {\bibinfo  {journal}
  {Mon. Not. Roy. Astron. Soc.}\ }\textbf {\bibinfo {volume} {413}},\ \bibinfo
  {pages} {628} (\bibinfo {year} {2011})}\BibitemShut {NoStop}%
\bibitem [{\citenamefont {Sugiyama}\ \emph {et~al.}(2017)\citenamefont
  {Sugiyama}, \citenamefont {Okumura},\ and\ \citenamefont {Spergel}}]{kSZ2}%
  \BibitemOpen
  \bibfield  {author} {\bibinfo {author} {\bibfnamefont {N.~S.}\ \bibnamefont
  {Sugiyama}}, \bibinfo {author} {\bibfnamefont {T.}~\bibnamefont {Okumura}}, \
  and\ \bibinfo {author} {\bibfnamefont {D.~N.}\ \bibnamefont {Spergel}},\
  }\href {\doibase 10.1088/1475-7516/2017/01/057} {\bibfield  {journal}
  {\bibinfo  {journal} {J. Cosmo. Astropart. Phys.}\ }\textbf {\bibinfo
  {volume} {01}},\ \bibinfo {pages} {057} (\bibinfo {year} {2017})}\BibitemShut
  {NoStop}%
\bibitem [{\citenamefont {{Fry}}(1984)}]{Frypap}%
  \BibitemOpen
  \bibfield  {author} {\bibinfo {author} {\bibfnamefont {J.~N.}\ \bibnamefont
  {{Fry}}},\ }\href {\doibase 10.1086/161913} {\bibfield  {journal} {\bibinfo
  {journal} {Astrophys. J}\ }\textbf {\bibinfo {volume} {279}},\ \bibinfo
  {pages} {499} (\bibinfo {year} {1984})}\BibitemShut {NoStop}%
\bibitem [{\citenamefont {{Gil-Mar{\'\i}n}}\ \emph {et~al.}(2012)\citenamefont
  {{Gil-Mar{\'\i}n}}, \citenamefont {{Wagner}}, \citenamefont {{Fragkoudi}},
  \citenamefont {{Jimenez}},\ and\ \citenamefont {{Verde}}}]{Gilmarin}%
  \BibitemOpen
  \bibfield  {author} {\bibinfo {author} {\bibfnamefont {H.}~\bibnamefont
  {{Gil-Mar{\'\i}n}}}, \bibinfo {author} {\bibfnamefont {C.}~\bibnamefont
  {{Wagner}}}, \bibinfo {author} {\bibfnamefont {F.}~\bibnamefont
  {{Fragkoudi}}}, \bibinfo {author} {\bibfnamefont {R.}~\bibnamefont
  {{Jimenez}}}, \ and\ \bibinfo {author} {\bibfnamefont {L.}~\bibnamefont
  {{Verde}}},\ }\href {\doibase 10.1088/1475-7516/2012/02/047} {\bibfield
  {journal} {\bibinfo  {journal} {J. Cosmo. Astropart. Phys.}\ }\textbf
  {\bibinfo {volume} {2012}},\ \bibinfo {eid} {047} (\bibinfo {year}
  {2012})}\BibitemShut {NoStop}%
\end{thebibliography}%

\end{document}